\documentclass[letterpaper,12pt]{article} 
\usepackage{osameet3} %% use version 3 for proper copyright statement
\usepackage{amsmath}
\usepackage{amsfonts}
\usepackage{amssymb}
\usepackage{amsthm}
\usepackage{amscd}
\usepackage[hidelinks]{hyperref}
\usepackage{siunitx}
\usepackage{graphicx}
\usepackage{booktabs}
\usepackage[ruled,linesnumbered]{algorithm2e}
\usepackage{multirow}
\usepackage{comment}
\usepackage{rotating}
\usepackage{tablefootnote}
\usepackage{threeparttable}
\usepackage{bm}
\usepackage{float}
\usepackage{setspace}
\usepackage{tabularx}

\usepackage{fancyhdr}
\begin{document}

\title{Tailoring Capture-Recapture Methods to Estimate Registry-Based Case Counts Based on Error-Prone Diagnostic Signals}

\author{Lin Ge, Yuzi Zhang, Kevin C. Ward, Timothy L. Lash, Lance A. Waller,  Robert H. Lyles}
\address{Department of Biostatistics and Bioinformatics, Emory University}
\email{lin.ge@emory.edu}
%%Uncomment the following line to override copyright year from the default current year.
%\copyrightyear{2022}

\onehalfspacing
\begin{abstract}
Surveillance research is of great importance for effective and efficient epidemiological monitoring of case counts and disease prevalence. Taking specific motivation from ongoing efforts to identify recurrent cases based on the Georgia Cancer Registry, we extend recently proposed “anchor stream” sampling design and estimation methodology. Our approach offers a more efficient and defensible alternative to traditional capture-recapture (CRC) methods by leveraging a relatively small random sample of participants whose recurrence status is obtained through a principled application of medical records abstraction. This sample is combined with one or more existing signaling data streams, which may yield data based on arbitrarily non-representative subsets of the full registry population. The key extension developed here accounts for the common problem of false positive or negative diagnostic signals from the existing data stream(s).  In particular, we show that the design only requires documentation of positive signals in these non-anchor surveillance streams, and permits valid estimation of the true case count based on an estimable positive predictive value (PPV) parameter. We borrow ideas from the multiple imputation paradigm to provide accompanying standard errors, and develop an adapted Bayesian credible interval approach that yields favorable frequentist coverage properties. We demonstrate the benefits of the proposed methods through simulation studies, and provide a data example targeting estimation of the breast cancer recurrence case count among Metro Atlanta area patients from the Georgia Cancer Registry-based Cancer Recurrence Information and Surveillance Program (CRISP) database.  
\end{abstract}

\doublespacing
%%%%%%%%%%%%%%%%%%%%%%%%%%%%%%%%%%%%%%%%%%%%%%%%%%%%%%%%%%%%%%%%
%% section 1: introduction
\section{Introduction}

\subsection{Background}

Capture-recapture (CRC) methods were originally developed for use in ecological studies designed to estimate wildlife populations living in a specific area  \cite{Petersen1896,Lincoln1930,Schnabel1938}. This approach involves capturing, tagging, releasing and recapturing target subjects, so that the total population size can be estimated based on the capture profiles of “caught” individuals who are identified in one or more capture efforts. There is a great deal of literature discussing the statistical methodologies behind such a sampling strategy  \cite{Borchers2002,Chao2001,Krebs1999}. In the traditional two-catch CRC setting, it is intuitive to create a two-by-two contingency table to describe the numbers of captured and recaptured subjects. More generally, there are  $2^T$ capture profiles created from the overlap of capture efforts when the number of capture efforts exceeds two ($T \geq 2$). Unfortunately, one profile is always unobservable because, by definition, it is comprised of individuals “caught” by none of the capture efforts. This missing value poses the fundamental inferential challenge for estimating the population size ($N$) based on CRC methods.

In epidemiological and public health-related research, applications of the CRC strategy to quantify vulnerable populations or case counts are becoming more common. CRC methods have been advocated for use in estimating case counts of many diseases and conditions, such as infant congenital syndrome \cite{Wittes1974,Hook1980}, the injection drug use \cite{Frischer1991,Platt2004,Hickman2004}, Huntington's disease \cite{Hook1992}, cancer \cite{McClish2004,Parkin2009}, and various infectious disease, such as HIV \cite{Poorolajal2017}, Hepatitis C \cite{Wu2005} and tuberculosis \cite{Dunbar2011,PerezDuque2020}.

When implementing CRC analysis based on real-world surveillance data, in general one must rely on empirically unverifiable modeling assumptions regarding the associations among the data streams to be used. For example, classical estimators for the $T = 2$ case \cite{Petersen1896, Lincoln1930, Chapman1951} assume that the two data streams operate independently at the population level. It is well known that such independence is often questionable in practice, and that flawed assumptions about dependency relationships among capture efforts can bias estimation of the population size \cite{Brenner1995, Seber1982}. As multiple sources \cite{Agresti1994,Cormack1999,Hook1995} have suggested, this challenge creates one of the shortcomings of popular CRC paradigms such as log-linear modeling often used in practice \cite{Baillargeon2007, Fienberg1972}. In light of this issue, some authors \cite{Chatterjee2016,Zhang2020,Zhang2022} have proposed variations on sensitivity analyses to acknowledge uncertainty about associations between CRC surveillance streams. 

One design-based way to ensure the necessary independence (sometimes known as Lincoln-Petersen, or “LP”) conditions in two-stream CRC is to conduct the second surveillance effort after the first, through principled random sampling of the target population (e.g., Seber et al. (1982) \cite{Seber1982}, Chao et al. (2008) \cite{Chao2008}). While this justifies the use of the classical Lincoln-Petersen or Chapman \cite{Chapman1951} estimator, a recent article \cite{Lyles2022} showed how a novel and more precise unbiased CRC estimator is unlocked in applications when the target population consists of a list or registry amenable to random sampling and ascertainment of disease status (providing a so-called “anchor stream” of representative data). In the absence of misclassification, this anchor stream approach can be used directly to calibrate data from one or more non-representative surveillance data streams toward efficient estimation of a cumulative case count among a registry population over a specified time period.

\subsection{Motivating Study and Methodological Challenge}

We draw specific motivation from the ongoing Cancer Recurrence Information and Surveillance Program (CRISP), which is augmenting the Georgia Cancer Registry (GCR) by developing the first US cancer recurrence registry of its kind. This project uses multiple data streams to signal possible cancer recurrences among GCR patients, and incorporates medical record abstraction by certified registrars for signal validation. True recurrences are defined as those that would be identified if subjected to full CRISP protocol-based review, among registry patients with initial primary non-metastatic tumors diagnosed on or after January 1st, 2013. While a full GCR registry population is being considered in CRISP, we focus here on a subset of the GCR breast cancer population for whom initial diagnosis and treatment occurred in one of five metro Atlanta hospitals and whose residential addresses have remained in the Atlanta area.  This latter criterion was intended to maximize the likelihood that cancer recurrence, were it to occur, would be captured in an Atlanta area facility. We utilize one existing surveillance data stream comprised of Georgia Commission on Cancer (CoC) recurrence reports, which has been signaling potential recurrences since 2013. On May 1st, 2022, we drew an anchor stream random sample of 200 breast cancer patients from the subset of the GCR, and CRISP abstractors conducted the necessary medical records review to ascertain whether each of those patients had experienced a true recurrence since their initial treatment date. Assuming the CoC recurrence reports were fully accurate, this design would unlock the original version of the anchor stream-based CRC estimator  \cite{Lyles2022}.

%While a full GCR registry population is being considered in CRISP, we focus here on a subset of the GCR breast cancer population whose residential addresses have remained in the Atlanta area and for whom initial diagnosis and treatment occurred in one of five metro Atlanta hospitals for which CRISP abstractors have access to electronic medical records.

In epidemiological surveillance, a common challenge is that disease status as signaled by existing data streams (e.g., CoC) can be prone to error. Such misclassification of the diagnostic assessment can subsequently lead to biased estimation when implementing CRC approaches. Relatively few articles have addressed this problem, particularly as it applies to human disease surveillance. Kendall et al. (2003) \cite{Kendall2003} and Conn and Cooch (2009) \cite{Conn2009} discussed  CRC with imperfect observations in the context of ecological studies, while Brenner (1996) \cite{Brenner1996} and Ramos et al. (2020) \cite{Ramos2020} considered error-prone surveillance data streams when false-positive and false-negative rates are assumed known. 

In this paper, we develop CRC methods to estimate epidemiological disease case counts accounting for misclassified diagnostic disease status. We utilize the anchor stream design \cite{Lyles2022}, for the dual purpose of ensuring independence between the two data streams and for estimating the unknown positive predictive value  (PPV) associated with error-prone signals from the existing non-anchor stream. These key objectives are accomplished by virtue of the fact that, to date, the anchor stream uses an arguably perfect (or “gold standard”) diagnostic approach, and consists of a random or representative sample of the population obtained subsequent to the signals from the existing stream. The anchor stream sample can be relatively small, identifying far fewer cases than the non-anchor stream. In the current work, we show how the design identifies the key PPV parameter required to justify the error-prone cell counts in the capture profiles, leading to an efficient case count estimator adjusted for misclassification. We show that the proposed approach extends readily to accommodate multiple non-anchor surveillance efforts; thus, it accommodates the general CRC setting $(T \geq 2)$, as long as one anchor stream sample is included.

%%%%%%%%%%%%%%%%%%%%%%%%%%%%%%%%%%%%%%%%%%%%%%%%%%%%%%%%%%%%%%%%
%% section 2: statistical methods
\section{Methods}

%. 2.1 Anchor stream, N_RS estimator
\subsection{Anchor Stream Design}
Lyles et al. (2022) \cite{Lyles2022} discuss the characteristics of an anchor stream sample, which can be feasible when surveilling disease within an enumerated registry population. For monitoring a cumulative case count among the registry over a specified follow-up period starting from an initial time point at which members are disease free, the key feature is that a simple or stratified random sample is drawn from the registry at the end of follow-up. Those selected are then assessed for whether disease has occurred, unless they were already identified as cases through one or more existing (though potentially non-representative) data streams.  The post-enumeration nature of the anchor stream sample keeps it “agnostic” of the existing streams, thus justifying classical CRC estimators based on the LP conditions \cite{Chao2008}. However, an anchor stream alone also provides its own valid and defensible sampling-based estimator, albeit often based on a much smaller sample size and number of identified cases than those obtained across the existing surveillance data stream(s). Assuming the total population size ($N_{tot}$) of the closed community or registry is known, the simple and familiar prevalence and case count estimators are:
\begin{align}
\hat{p}_{RS} = n^+/n,~\hat{N}_{RS} = N_{tot}\hat{p}_{RS},~\hat{Var}(\hat{N}_{RS}) = N_{tot}^2\hat{Var}(\hat{p}_{RS}), \label{N_RS}
\end{align}
where $n^+$ denotes the number of test positives and $n$ denotes the number of individuals selected into the anchor stream. Given that the population is closed and finite, the estimated variance of $\hat{p}_{RS}$ incorporates a finite population correction (FPC) given by Cochran (1977) \cite{Cochran77}, i.e.,
\begin{align}
\hat{Var}(\hat{p}_{RS}) = [\frac{n(N_{tot} - n)}{N_{tot}(n-1)}]\frac{\hat{p}_{RS}(1-\hat{p}_{RS})}{n}.
\end{align}
When the sample size $n$ of the anchor stream is large relative to the population size $N_{tot}$, the effect of the FPC is well known to be substantial.

Assuming that disease assessments conducted among both the anchor stream sample and via an existing non-representative surveillance system that identifies cases over the follow-up period are accurate, an unbiased estimator for the case count that will often be much more precise than either the sampling-based estimator in (\ref{N_RS}) or the classical CRC estimator (e.g., Chapman’s) unlocked by design is the following:
\begin{align}\label{original_MLE}
    &\hat{N}_{Anchor} = n_{11} + n_{10} + n_{01}/\psi,\nonumber\\
    &\hat{Var}(\hat{N}_{Anchor}) = n_{01}(1-\psi)/\psi^2
\end{align}
where $\psi$ is the selection probability into the anchor stream and $n_{11}$, $n_{10}$, and $n_{01}$ are the numbers of cases identified by both streams, by the existing but not the anchor, and by the anchor but not the existing stream, respectively \cite{Chen2020,Lyles2021a}. Variations on the estimator in (\ref{original_MLE}) and inferential considerations were subsequently proposed \cite{Lyles2022}. While the anchor stream must satisfy key design requirements, it can yield much greater precision than classical CRC estimators even based on a small anchor stream random sample that augments an existing surveillance system.

%. 2.2 add data observation
\subsection{Plug-in Estimator with Known PPV} \label{section_simple_plugin_approach}

We label the existing surveillance effort, which typically selects those at high risk of disease preferentially, as Stream 1. The anchor stream, implemented by design to collect a random sample for testing independently of Stream 1, is labeled as Stream 2. In practice, the misclassification of disease status is a common problem in Stream 1. For an extension of the original anchor stream estimation process to accommodate error-prone cell counts, we assume that an accurate diagnostic method (e.g., protocol-based medical records abstraction) is employed to assess disease status in Stream 2.

\begin{table}[ht]
    \centering
    \caption{Observations for Two-stream Capture-Recapture Analysis}    \label{Obs_type0}
    \begin{threeparttable}[b]
    \begin{tabular}{|c|c|c|c|}
    \hline
     &\multicolumn{2}{c|}{\# of Cases Found in 2nd stream}  &  \\
    \hline
     \# of Cases Found in 1st stream & \hspace*{10mm} Yes \hspace*{10mm} &\hspace*{10mm}  No  \hspace*{10mm} & \hspace*{2mm} Total \hspace*{2mm}\\
    \hline
     Yes & $m_{11}$ &  $m_{10}$ & $m_{1.}$\\
     \hline
     No \tnote{*}  &   $m_{01}$ &  $m_{00} =?$ & $m_{.1}$ \\
     \hline
    \end{tabular}
     \begin{tablenotes}
        \item[*] Cases not found in a given stream are either unsampled in that stream, or are sampled and tested negative.
  \end{tablenotes}
    \end{threeparttable}
\end{table}
The observed cell counts $m_{11}$, $m_{10}$, $m_{01}$ in Table \ref{Obs_type0} may vary from the corresponding ``true" counts ($n_{11}$, $n_{10}$, and $n_{01}$) when Stream 1 employs an imperfect diagnostic method. We note that false negatives surveilled in Stream 1 are the same as missed cases in the CRC paradigm, so that with an accurate test employed in Stream 2, it is only necessary to adjust for potential false positives in the $m_{10}$ cell. To make this adjustment, we define the positive predictive value in Stream 1 as follows: $PPV_1$ =  Pr(true positive $|$ sampled in 1 and tested positive). Assuming for the moment that this parameter is known, one can correct for misclassification of the observed case counts in Stream 1 using the following extension of the original anchor stream-based CRC estimator in (\ref{original_MLE}):
\begin{equation}
    \hat{N}_{{\psi}}={PPV}_{1}(m_{11} +m_{10}) + \frac{m_{01}}{{\psi}}  \label{MLE_psi}
\end{equation}
Here, $\psi$ is the probability of identifying a case in Stream 2 given the case is missed in Stream 1, which is efficiently estimated as $\hat{\psi} = n/N_{tot}$, i.e., the overall sampling rate into Stream 2.

We propose a variance estimator to accompany (\ref{MLE_psi}) by leveraging the multiple imputation (MI) paradigm \cite{Rubin1987}, yielding an adjustment for the extra uncertainty in the number of cases that would have been identified in Stream 1 in the absence of misclassification. In the MI method, we consider this count as a missing value and impute it as a random variate generated as $Binomial(m_{11}+m_{10},~{PPV}_{1})$. Then the variance is calculated from a summation of the within- and between-imputation components of variance. The details of steps to calculate this variance are provided in Appendix \ref{appendix_A}. However, in practice, the parameter $PPV_1$ will generally be unknown and must instead be estimated from the data. We address this more typical scenario in subsequent sections.

%. 2.3 ML approach
\subsection{Maximum Likelihood (ML) Approach}

In this section, an alternative estimator is derived via maximum likelihood (ML). As before, the approach only requires documentation of positive signals in the non-anchor surveillance stream. However, here we obtain valid estimation of the true case count based on an estimable positive predictive value parameter using all of the observed data. This estimator is the MLE for $N$ under a general multinomial model for the six cell counts defined in Table \ref{collapse_table_with_prob}.

\begin{table}[ht]
\centering
\caption{Cell Counts and Likelihood Contributions for Collapsed Table Observations from the Two-stream Disease Surveillance}\label{collapse_table_with_prob}
\begin{threeparttable}[b]
 \begin{tabular}{|c|l|l|l|} 
 \hline
  Cell & \multicolumn{1}{c|}{Observation Type} & Likelihood Contribution & Re-Parameterization \tnote{ }\\ 
 \hline
 \multirow{2}{*}{$n_1$} & Test + in Stream 2, &\multirow{2}{*}{$\psi PPV_1 \pi^*_{s1}\phi$}  &\multirow{2}{*}{$\psi PPV_1 \theta$}\\
  & Test + in Stream 1  &  &\\ \hline
    \multirow{2}{*}{$n_2$} & Test + in Stream 2, & $\psi(1-NPV_1)(1-\pi^*_{s1})\phi +$ &\multirow{2}{*}{$\psi (\pi - \theta)$}\\\
  & Not test + in Stream 1 & $\psi\pi_{\bar{s}1}(1-\phi)$ &\\\hline
  \multirow{2}{*}{$n_3$} & Test \textminus ~ in Stream 2, &\multirow{2}{*}{$\psi(1-PPV_1)\pi^*_{s1}\phi$}  &\multirow{2}{*}{$\psi (1-PPV_1) \theta$}\\
  & Test + in Stream 1  &  &\\ \hline
\multirow{2}{*}{$n_4$} & Test \textminus~in Stream 2, & $\psi NPV_1 (1-\pi^*_{s1})\phi +$ &\multirow{2}{*}{$\psi (1 -\pi )$}\\
  & Not test + in Stream 1  & $\psi(1-\pi_{\bar{s}1})(1-\phi)$ &\\ \hline
  \multirow{2}{*}{$n_5$} & Not sampled in Stream 2, &\multirow{2}{*}{$(1-\psi)\pi^*_{s1}\phi$}  &\multirow{2}{*}{$(1-\psi)\theta$}\\
  & Test + in Stream 1  &  &\\ \hline
   \multirow{2}{*}{$n_6$} &  Not sampled in Stream 2, & $(1-\psi)(1-\pi^*_{s1})\phi +$ &\multirow{2}{*}{$(1-\psi) (1 - \theta)$}\\
  & Not test + in Stream 1  & $(1-\psi)(1-\phi)$ &\\
  \hline
 \end{tabular}
    \begin{tablenotes}
        \item[1] Re-parameterization $\theta = \pi^*_{s1}\phi$, $\pi =  [1-NPV_1(1-\pi^*_{s1})]\phi + \pi_{\bar{s}1}(1-\phi)$.
        \item[2] Not test + in Stream 1 indicates test negative or not sampled in Stream 1.
  \end{tablenotes}
 \end{threeparttable}
\end{table}

The likelihood contributions given in Table \ref{collapse_table_with_prob} are based on defining the parameters, $\phi=$ Pr(Sampled in Stream 1), $\psi=$ Pr(Sampled in Stream 2), $\pi^*_{s1}=$ Pr(Test + $|$ Sampled in 1), $\pi_{\bar{s}1}=$ Pr(True + $|$ Sampled not in 1) as well as the positive predictive value $PPV_1$ (defined previously) and negative predictive values, $NPV_1=$ Pr(True - $|$ Sampled in 1 and test -). Letting $p_j$ denote the likelihood contribution corresponding to the $j$th cell, the vector of six cell counts can be modeled as a multinominal sample, with likelihood proportional to $\prod_{j = 1}^6 p_j^{n_j}$, i.e., 
\begin{equation}
(n_1,n_2,\cdots,n_6)\sim~multinomial (N_{tot}; p_1, p_2, \cdots, p_6) \nonumber
\end{equation}

Not all of the parameters in the third column of Table \ref{collapse_table_with_prob} are identifiable, due to the fact that negative test results are assumed unavailable in the non-anchor surveillance data stream. Therefore, we define two new parameters ($\theta$ and $\pi$) by grouping some of the non-identifiable parameters together. Importantly, the key parameters ($\phi$, $\theta$, $\pi$, $PPV_1$) for defining the overall disease prevalence calculation are identifiable. The re-parameterization yields the MLEs for the following new parameters.

\begin{itemize}
    \item $\psi$: Pr(Sampled in Stream 2), $\hat{\psi} = \frac{n_1+n_2+n_3+n_4}{N_{tot}}$
    \item $\theta = \pi^*_{s1}\phi$, $\hat{\theta} = \frac{n_1+n_3 +n_5}{N_{tot}}$
    \item $\pi = [1-NPV_1(1-\pi^*_{s1})]\phi + \pi_{\bar{s}1}(1-\phi)$,  $\hat{\pi} = \frac{1}{N_{tot}}[(n_1+n_3+n_5) + \frac{n_2}{n_2+n_4}(n_2+n_4 + n_6)]$
    \item $PPV_1$: Pr(True + $|$ Sampled and Test + in 1), $\hat{PPV}_1 = \frac{n_1}{n_1+n_3}$
\end{itemize}

The overall disease prevalence is a function of these parameters, i.e. $PPV_1\theta + \pi-\theta$ (see Appendix \ref{appendix_B}); as a result, an intuitive form for the ML estimator for $N$ after re-parameterization in Table \ref{collapse_table_with_prob} is as follows.
\begin{align}\label{MLE_collapse}
    \hat{N}_{\hat{\psi}^{*}} &= \hat{PPV}_1(m_{11} + m_{10}) + \frac{m_{01}}{\hat{\psi}^{*}} 
    %&=  m_{11} + m_{10}\hat{PPV}_{10} + \frac{m_{01}}{\hat{\psi}^{**}} %\\
    %or~\Tilde{N}_{\hat{\psi}^{**}} &=
    %\hat{PPV}_1(m_{11} + m_{10}) + m_{01}/\hat{\psi}^{**} \nonumber\\
    %& = m_{11} + m_{10}\hat{PPV}_{10} + m_{01}/\hat{\psi}^{**} \label{MLE_collapse2}
\end{align}
where $\hat{\psi}^{*} = \frac{n_2+n_4}{n_2 + n_4 + n_6}$, and $m_{11} = n_1$, $m_{10} = n_3+n_5$, $m_{01} = n_2 $ are the observed cell counts from Table \ref{Obs_type0}. The derivation for the MLE is given in Appendix \ref{appendix_B}, and details to facilitate estimation of its variance can be found in Appendix \ref{appendix_A}.

With regard to the standard errors (SEs) in empirical studies, the ML theory-based variance estimator accompanying (\ref{MLE_collapse}) based on the multinomial model for the cell counts in Table \ref{collapse_table_with_prob} results in overly conservative 95\% Wald-type confidence intervals when the sampling rate into the anchor stream becomes moderate to large. This is because increasing the sampling rate into the anchor stream creates finite population sampling-related adjustments to the standard multinomial variance-covariance matrix that are difficult to explicitly quantify given the combination of arbitrarily non-random sampling in Stream 1 with random sampling in Stream 2.  To better approximate the SE, we substitute a second version of an MI paradigm-based variance estimator that accounts for uncertainty in the $PPV_1$ parameter (see Appendix \ref{appendix_A}). This improves the SE estimate as well as overall Wald-type CI coverage. However, in the next section, we provide another approach to interval estimation with appealing properties.

%. 2.4 Bayesian Credible Interval
\subsection{An Adapted Bayesian Credible Interval Approach for Inference}

A broad CRC literature considers applying normalizing transformations to improve Wald-type CI coverage, such as the log transformation  \cite{Chao1987}, the inverse transformation \cite{Jensen1989,Krebs1999} and a transformed logit confidence interval \cite{Sadinle2009}. Here, we adapt a Bayesian credible interval \cite{Brown2001,Sangeetha2013,Lyles2022} based on a weakly informative Dirichlet prior on multinomial cell probabilities, to improve the coverage of intervals accompanying the ML estimator in (\ref{MLE_collapse}).

%, all of which share the same form of the estimator for $N$ as 
%\begin{equation}
%    \hat{N}_{\Tilde{\psi}} = \hat{PPV}_{1}(m_{11} + m_{10}) + \frac{m_{01}}{\Tilde{\psi}}
%\end{equation} 

This proposed approach involves two main steps to construct the Bayesian credible interval (see details in Appendix \ref{appendix_D}). First, we adjust the false positive case count impacting the $m_{10}$ of Table \ref{Obs_type0} by estimable posterior draws [$\Tilde{PPV}_{10,s}^*, (s = 1, \cdots, S)$] corresponding to a new PPV parameter specific to that cell (see Appendix \ref{appendix_C} for the relationship between $PPV_1$ and $PPV_{10}$). These draws leverage a Dirichlet posterior distribution based on a Jeffreys' $Dirichlet$(0.5, 0.5, 0.5) prior. Second, for each posterior draw $s$, a further set of the posterior samples for the vector of accurate cell counts $(\Tilde{m}_{11,(s,t)}^*,\Tilde{m}_{10,(s,t)}^*,\Tilde{m}_{01,(s,t)}^*), (s = 1, \cdots S; t = 1, \cdots T)$ is generated to mimic posterior samples of the estimated $\hat{N}_{\hat{\psi}^*,(s,t)}^*$. An initial version of a Bayesian credible interval that we refer to as the unadjusted credible interval, or $(LL_{unadj}, UL_{unadj})$, is based on a (2.5\%, 97.5\%) percentile interval formed from these posterior estimates. This unadjusted interval is recommended in the case of a relatively small estimated prevalence (less than 20\%). A second version of a credible interval, $(LL_{adj}, UL_{adj})$, applies scales and shift adjustments to the unadjusted credible interval and is recommended for high prevalence (over 20\%) scenarios (see Appendix \ref{appendix_D}).

%. 2.5 Extension the Two-stream CRC Analysis to Multiple-stream Case
\subsection{Extension the Two-stream CRC Analysis to Multiple-stream Case}

Lyles et al. (2022) \cite{Lyles2022} discuss a straightforward extension of the two-stream anchor stream-based CRC analysis to accommodate multiple non-anchor streams when an accurate diagnostic method is applied in each. Extending this approach to account for misclassification, we assume there are multiple data streams $T(>2)$ used for surveillance and that one of these (Stream $T$) is an anchor stream by design that is independent (agnostic) of the other streams and uses an accurate diagnostic test. Then regardless of the dependencies among the $T-1$ non-anchor streams and regardless of their tendency to misdiagnose cases, a valid estimator results from creating a single composite non-anchor stream consisting of all signaled cases identified by at least one of the original $T-1$ streams. This composite stream takes the role of Stream 1, while anchor Stream $T$ takes the role of Stream 2. It is then straightforward to apply the ML estimator (\ref{MLE_collapse}) directly, where a compound positive predictive value is estimated for the composite Stream 1.

%%%%%%%%%%%%%%%%%%%%%%%%%%%%%%%%%%%%%%%%%%%%%%%%%%%%%%%%%%%%%%%%
%% section 3: simulation study

\section{Simulation Study}

In this section, we present simulation studies designed to compare the performances of estimators of $N$ proposed in the previous section. In the first simulation setting, we simulated the data based on a hypothetical scenario mimicking breast cancer recurrence surveillance data in the Georgia Cancer Registry. We first generated a population of size $N_{tot} = 1000$ and created two data streams for CRC analysis. Stream 1 represents a non-random surveillance effort with error-prone disease status (i.e., subject to false positive signals), and Stream 2 is an anchor stream that adopts perfect testing and operates independently of Stream 1. We separate the total population into two strata (40\% vs 60\%), based on a binary characteristic (e.g., representing symptom status or cancer stage). Informed by preliminary data, we set the prevalence in stratum 1 (40\%) and stratum 2 (60\%) to be 0.182 and 0.046 respectively, which results in the number of true disease cases fixed at 100 in each simulated dataset; i.e., the overall prevalence is 10\%. For Stream 1, the probabilities of selection for testing in stratum 1 and 2 were set to 0.695 and 0.347 (thus, Stream 1 oversamples high-risk subjects with symptoms or higher cancer stage). Moreover, we set the sensitivity of the testing approach in Stream 1 to 0.9 and the specificity to 0.95 (corresponding to $PPV_1 = 0.72$). We then varied the sampling rate into Stream 2, i.e., $p_2$ = (5\%, 10\%, 20\%, 50\%), so that the corresponding sample sizes are (50, 100, 200, 500). Only positive records are documented in Stream 1 for analysis, while both positive and negative results are incorporated from Stream 2. Table \ref{simu_results1a} summarizes the results of this initial simulation study based on 2000 simulation runs per scenario to compare estimators of the case count.

\begin{table}[ht]
\centering
\caption{Simulation Study 1 to Compare the Performance of Estimators, $N_{tot} = 1000$ and $N = 100$ Diseased Subjects}\label{simu_results1a}
 \scalebox{1.0}{
\begin{threeparttable}[b]
\begin{tabular}{l|c|ccccccccc}
  \hline
Settings &Estimators & $\hat{N}$ & SD &  Avg.SE & CI Coverage (\%) \tnote{a} & Width  \\ 
  \hline
& $\hat{N}_{Chap}$ & 132.2 & 54.9 & 41.7 & 78.9 & 912.2 \\ 
  $p_2= 5 \% $ & $\hat{N}_{RS}$ & 99.4 & 42.4 & 40.2 & 90.5 & 157.4 \\ 
  $n_2= 50 $ & $\hat{N}_{{\psi}}$ & 98.7 & 29.5 & 27.3 & 88.4 & 107.0 \\ 
  & $\hat{N}_{\hat{\psi}^{*}}$ & 99.5 & 34.9 & 32.0 &88.3 (\textbf{95.3}) & 125.3 (\textbf{133.3}) \\ 
   \hline
& $\hat{N}_{Chap}$ & 138.2 & 42.6 & 35.7 & 58.4 & 304.5 \\ 
  $p_2= 10 \% $ & $\hat{N}_{RS}$ & 100.0 & 28.6 & 28.2 & 93.0 & 110.7 \\ 
  $n_2= 100 $ & $\hat{N}_{{\psi}}$ & 99.7 & 19.9 & 20.2 & 91.7 & 79.1 \\ 
  & $\hat{N}_{\hat{\psi}^{*}}$ & 99.4 & 23.6 & 22.9 & 92.9 (\textbf{95.9}) & 89.9 (\textbf{91.6}) \\ 
   \hline
& $\hat{N}_{Chap}$ & 139.8 & 30.0 & 26.1 & 31.6 & 141.6 \\ 
  $p_2= 20 \% $ & $\hat{N}_{RS}$ & 99.1 & 19.3 & 18.8 & 93.8 & 73.8 \\ 
  $n_2= 200 $ & $\hat{N}_{{\psi}}$ & 99.9 & 13.6 & 14.0 & 93.7 & 54.7 \\ 
  & $\hat{N}_{\hat{\psi}^{*}}$ & 99.6 & 15.8 & 15.8 & 93.9 (\textbf{94.8}) & 62.3 (\textbf{61.2}) \\ 
   \hline
& $\hat{N}_{Chap}$ & 139.3 & 16.0 & 14.3 & 3.4 & 64.5 \\ 
  $p_2= 50 \% $ & $\hat{N}_{RS}$ & 100.0 & 9.5 & 9.5 & 95.0 & 37.2 \\ 
  $n_2= 500 $ & $\hat{N}_{{\psi}}$ & 100.1 & 7.5 & 7.8 & 95.8 & 30.6 \\ 
  & $\hat{N}_{\hat{\psi}^{*}}$ & 99.9 & 7.8 & 8.7 & 97.2 (\textbf{95.8}) & 34.5 (\textbf{31.4}) \\ 
   \hline
\end{tabular}
\begin{tablenotes}
    \item[$a$] Wald-based CIs are evaluated, except we report the transformed logit CI \cite{Sadinle2009} for $\hat{N}_{Chap}$ and the Bayesian credible interval (from 10,000 Dirichlet posterior draws per simulation) for $\hat{N}_{\hat{\psi}^*}$ in (\textbf{bold}).
    %\item[$b$] The standard error of $\hat{N}_{\hat{\psi}^{*}}$ is calculated from a MI paradigm-based variance estimator (see Appendix A), instead of the delta method approach.
\end{tablenotes}
\end{threeparttable}
}
\end{table}

In this simulation study, we compare the naive Chapman estimator $\hat{N}_{Chap}$ without justifying any misclassified cell counts, to all estimators discussed in section 2. As expected, only $\hat{N}_{Chap}$ yields biased estimation (due to the misclassification), while the others yield unbiased estimates. In each setting, the anchor stream-based CRC estimator $\hat{N}_{{\psi}}$ or $\hat{N}_{\hat{\psi}^{*}}$, yields smaller empirical standard deviation (SD) than the sampling-based estimator $\hat{N}_{RS}$. Additionally, the justification by the positive predictive value is highly effective, whereby the plug-in estimator $\hat{N}_{{\psi}}$ utilizes the true $PPV_1$, while the ML estimator $\hat{N}_{\hat{\psi}^{*}}$ applies the maximum likelihood estimator for $PPV_1$. Beacause $PPV_1$ is estimated, $\hat{N}_{\hat{\psi}^{*}}$ loses some efficiency compared to the plug-in estimator; however, we note again that it is the more realistic estimator in real data applications.

Wald-type CIs accompanying the proposed estimators tend to be anti-conservative when the sampling rate into Stream 2 is small. In comparison, the proposed adapted Bayesian credible interval effectively propagates uncertainty in the ML estimators and shows stable and well-behaved interval estimation across the full range of sampling rate $(p_2)$ examined in Table \ref{simu_results1a}.

We designed simulation 2 by enlarging the total population from 1000 to 5000. With the hypothetical disease surveillance (Stream 1) conducted under the same conditions as in simulation 1, the true case number in this setting is 500. We then examined a different set of sampling rates (2\%, 4\%, 10\%, 20\%) for the anchor stream. The corresponding sample sizes in stream 2 are (100, 200, 500, 1000), and Table \ref{simu_results1c} shows the results. We note that all proposed estimators continue to perform well and the adapted credible interval provides excellent coverage, even when Stream 2 consists of only a 2\% random sample of the population.

\begin{table}[ht]
\centering
\caption{Simulation Study 2 to Compare the Performance of Estimators, from $N_{tot} = 5000$ and $N = 500$ Diseased Subjects}\label{simu_results1c}
 \scalebox{1.0}{
\begin{threeparttable}[b]
\begin{tabular}{l|c|ccccccccc}
  \hline
Settings &Estimators & $\hat{N}$ & SD &  Avg.SE & CI Coverage (\%) \tnote{a} & Width  \\ 
  \hline
& $\hat{N}_{Chap}$ & 696.8 & 229.3 & 188.0 & 63.7 & 1657.8 \\ 
  $p_2= 2 \% $ & $\hat{N}_{RS}$ & 496.0 & 148.1 & 146.4 & 92.8 & 573.7 \\ 
  $n_2= 100 $ & $\hat{N}_{{\psi}}$ & 499.2 & 104.1 & 103.5 & 91.4 & 405.9 \\ 
  & $\hat{N}_{\hat{\psi}^{*}}$ & 501.1 & 122.7 & 118.7 & 92.7 (\textbf{95.7}) & 465.3 (\textbf{482.4}) \\ 
   \hline
& $\hat{N}_{Chap}$ & 698.8 & 157.0 & 137.4 & 39.0 & 729.2 \\ 
  $p_2= 4 \% $ & $\hat{N}_{RS}$ & 502.0 & 106.4 & 103.6 & 93.4 & 406.1 \\ 
  $n_2= 200 $ & $\hat{N}_{{\psi}}$ & 502.0 & 75.8 & 74.1 & 93.1 & 290.7 \\ 
  & $\hat{N}_{\hat{\psi}^{*}}$ & 501.8 & 83.4 & 84.2 & 94.3 (\textbf{94.8}) & 337.9 (\textbf{337.4}) \\ 
   \hline
& $\hat{N}_{Chap}$ & 698.7 & 91.2 & 86.6 & 7.0 & 381.3 \\ 
  $p_2= 10 \% $ & $\hat{N}_{RS}$ & 501.3 & 62.5 & 63.6 & 94.8 & 249.3 \\ 
  $n_2= 500 $ & $\hat{N}_{{\psi}}$ & 501.5 & 44.9 & 46.2 & 95.2 & 181.2 \\ 
  & $\hat{N}_{\hat{\psi}^{*}}$ & 501.1 & 52.0 & 52.8 & 94.9 (\textbf{95.5}) & 206.1 (\textbf{206.4}) \\ 
   \hline
 & $\hat{N}_{Chap}$ & 698.8 & 62.3 & 59.3 & 0.5 & 246.8 \\ 
  $p_2= 20 \% $ & $\hat{N}_{RS}$ & 500.4 & 42.2 & 42.4 & 96.0 & 166.2 \\ 
  $n_2= 1000 $ & $\hat{N}_{{\psi}}$ & 500.6 & 31.0 & 31.5 & 95.3 & 123.6 \\ 
  & $\hat{N}_{\hat{\psi}^{*}}$ & 499.8 & 33.9 & 36.0 & 96.0 (\textbf{96.4}) & 141.0 (\textbf{138.4}) \\ 
   \hline
\end{tabular}
\begin{tablenotes}
    \item[$a$] Wald-based CIs are evaluated, except we report the transformed logit CI \cite{Sadinle2009} for $\hat{N}_{Chap}$ and the Bayesian credible interval (from 10,000 Dirichlet posterior draws per simulation) for $\hat{N}_{\hat{\psi}^*}$ in (\textbf{bold}).
    %\item[$b$] The standard error of $\hat{N}_{\hat{\psi}^{*}}$ is calculated from a MI paradigm-based variance estimator (see Appendix A), instead of the delta method approach.
\end{tablenotes}
\end{threeparttable}
}
\end{table}

Simulation study 3 aims to assess the performance of estimators when misclassification of disease status causes a major impact, i.e., when the sensitivity and specificity (and correspondingly the $PPV_1$) of surveillance Stream 1 are relatively lower than might typically be expected. Here, we keep all parameters the same as in simulation 1, except that the sensitivity and specificity take values of (0.90, 0.95), (0.70, 0.95), (0.90, 0.80), (0.70, 0.80), corresponding to $PPV_1$ values of 0.72, 0.66, 0.39, 0.33. The sampling rate into Stream 2 is set to be 10\%, and the results of this setting are presented in Table \ref{simu_results3}.

\begin{table}[ht]
\centering
\caption{Simulation Study 3 to Evaluate the Performance of Estimators When the Error-prone Disease Status Causes a Major Impact}\label{simu_results3}
\scalebox{1.0}{
\begin{threeparttable}[b]
\begin{tabular}{c|c|ccccccc}
  \hline
Settings &Estimators & $\hat{N}$ & SD &  Avg.SE & CI Coverage (\%) \tnote{a} & Width  \\ 
  \hline
  $Se = 0.90$ & $\hat{N}_{Chap}$ & 138.2 & 42.6 & 35.7 & 58.4 & 304.5 \\ 
  $Sp = 0.95$ & $\hat{N}_{RS}$ & 100.0 & 28.6 & 28.2 & 93.0 & 110.7 \\ 
  $(PPV_1 = 0.72)$ & $\hat{N}_{{\psi}}$ & 99.7 & 19.9 & 20.2 & 91.7 & 79.1 \\ 
  & $\hat{N}_{\hat{\psi}^{*}}$ & 99.4 & 23.6 & 22.9 & 92.9 (\textbf{95.9}) & 89.9 (\textbf{91.6}) \\ 
   \hline
  $Se = 0.70$ & $\hat{N}_{Chap}$ & 148.7 & 61.2 & 48.6 & 68.4 & 505.1 \\
  $Sp = 0.95$ & $\hat{N}_{RS}$ & 99.8 & 28.7 & 28.2 & 93.1 & 110.4 \\
  $(PPV_1 = 0.66)$ & $\hat{N}_{{\psi}}$ & 99.8 & 23.0 & 22.7 & 91.3 & 89.0 \\
  & $\hat{N}_{\hat{\psi}^{*}}$ & 99.5 & 25.7 & 24.9 & 92.6 (\textbf{95.4}) & 97.6 (\textbf{98.2}) \\
   \hline
$Se = 0.90$ & $\hat{N}_{Chap}$ & 254.9 & 81.5 & 66.5 & 0.0 & 527.9 \\
 $Sp = 0.80$ & $\hat{N}_{RS}$ & 100.6 & 28.5 & 28.3 & 93.0 & 110.9 \\
 $(PPV_1 = 0.39)$ & $\hat{N}_{{\psi}}$ & 100.0 & 20.7 & 20.7 & 92.0 & 81.0 \\
  & $\hat{N}_{\hat{\psi}^{*}}$ & 100.2 & 27.1 & 26.2 & 92.4 (\textbf{95.3}) & 101.2 (\textbf{102.6}) \\
   \hline
$Se = 0.70$ & $\hat{N}_{Chap}$ & 303.4 & 132.9 & 102.5 & 0.0 & 1203.0 \\
$Sp = 0.80$ & $\hat{N}_{RS}$ & 100.3 & 28.4 & 28.2 & 93.2 & 110.7 \\
$(PPV_1 = 0.33)$ & $\hat{N}_{{\psi}}$ & 100.5 & 22.2 & 23.2 & 93.0 & 90.9 \\
  & $\hat{N}_{\hat{\psi}^{*}}$ & 100.1 & 28.1 & 27.7 & 93.6 (\textbf{95.2}) & 106.8 (\textbf{107.3})\\
   \hline
\end{tabular}
\begin{tablenotes}
    \item[$a$] Wald-based CIs are evaluated, except we report the transformed logit CI \cite{Sadinle2009} for $\hat{N}_{Chap}$ and the Bayesian credible interval (from 10,000 Dirichlet posterior draws per simulation) for $\hat{N}_{\hat{\psi}^*}$in (\textbf{bold}).
    %\item[$b$] The standard error of $\hat{N}_{\hat{\psi}^{*}}$ is calculated from a MI paradigm-based variance estimator (see Appendix A), instead of the delta method approach.
\end{tablenotes}
\end{threeparttable}
}
\end{table}

Table \ref{simu_results3} summarizes the performance of competing estimators as the severity of misclassification due to error-prone disease diagnostic signals increases. As expected, all proposed estimators remain unbiased regardless of the increase in misclassification from Stream 1. We note here that the SD of the ML estimator $\hat{N}_{\hat{\psi}^*}$ is close to that of $\hat{N}_{RS}$ when the $PPV_1$ becomes very small. This is because the CRC analysis depends almost exclusively on the anchor stream alone when accurate test information from stream 1 becomes more and more scarce. This simulated setting suggests that the proposed CRC estimators remain valid in all cases considered, but that the gain in estimation efficiency relative to the estimator $\hat{N}_{RS}$ tends to decrease as $PPV_1$ deteriorates. 

Finally, simulation 4 was designed to further examine scenarios when the true prevalence becomes larger. We kept the population size $N_{tot} = 1000$, but varied the disease prevalence (10\%, 30\%, 50\%). Stream 1 was drawn under the same conditions as before, but we consider three different sampling rates,  (10\%, 30\%, 50\%) for the anchor stream. We compare the performance of the proposed ML estimator $\hat{N}_{\hat{\psi}^{*}}$ with the random sampling-based estimator $\hat{N}_{RS}$ in table \ref{simu_results4}.

\begin{table}[ht]
\centering
\caption{Simulation Study 4 to Compare the Performance of Estimators under Different Prevalence Levels}\label{simu_results4}
 \scalebox{1.0}{
\begin{threeparttable}[b]
\begin{tabular}{l|c|c|ccccccccc}
  \hline
Prevalence & $p_2$ &Estimators & $\hat{N}$ & SD &  Avg.SE & CI Coverage (\%) \tnote{a} & Width  \\ 
  \hline
& \multirow{2}{*}{10\%} & $\hat{N}_{RS}$  & 99.7 & 28.8 & 28.3 & 92.7 & 110.8 \\ 
  $p = 10 \% $ & &$\hat{N}_{\hat{\psi}^{*}}$ & 99.9 & 23.8 & 22.9 & 92.8 (\textbf{94.8}) & 89.8 (\textbf{91.9})\\ 
 & \multirow{2}{*}{30\%}& $\hat{N}_{RS}$ & 99.5 & 14.5 & 14.5 & 94.1 & 56.7 \\ 
  $N_{true}= 100 $ && $\hat{N}_{\hat{\psi}^{*}}$ & 99.7 & 11.8 & 12.4 & 95.4 (\textbf{95.2}) & 48.5 (\textbf{47.1}) \\ 
    & \multirow{2}{*}{50\%}& $\hat{N}_{RS}$ & 100.2 & 9.5 & 9.5 & 94.9 & 37.3 \\ 
  && $\hat{N}_{\hat{\psi}^{*}}$ & 100.2 & 7.7 & 8.7 & 97.1 (\textbf{95.8}) & 33.9 (\textbf{31.5}) \\ 
   \hline
   & \multirow{2}{*}{10\%} & $\hat{N}_{RS}$  & 300.1 & 43.9 & 43.6 & 94.7 & 170.9 \\ 
  $p = 30 \% $ & &$\hat{N}_{\hat{\psi}^{*}}$ & 300.4 & 34.2 & 37.4 & 96.4 (\textbf{94.2}) & 146.5 (\textbf{131.2})\\ 
 & \multirow{2}{*}{30\%}& $\hat{N}_{RS}$ & 299.7 & 21.9 & 22.2 & 95.4 & 86.8 \\ 
  $N_{true}= 300 $ && $\hat{N}_{\hat{\psi}^{*}}$ & 299.9 & 17.3 & 19.3 & 97.0 (\textbf{94.7}) & 75.5 (\textbf{67.3}) \\ 
    & \multirow{2}{*}{50\%}& $\hat{N}_{RS}$ & 300.1 & 14.7 & 14.5 & 94.7 & 56.9 \\ 
  && $\hat{N}_{\hat{\psi}^{*}}$ & 300.0 & 11.4 & 12.9 & 97.2 (\textbf{94.6}) & 50.6 (\textbf{44.4}) \\ 
   \hline
   & \multirow{2}{*}{10\%} & $\hat{N}_{RS}$  & 499.6 & 48.0 & 47.7 & 94.5 & 186.9 \\ 
  $p = 50 \% $ & &$\hat{N}_{\hat{\psi}^{*}}$ & 500.1 & 38.8 & 46.9 & 97.9 (\textbf{94.7}) & 184.0 (\textbf{154.1})\\ 
 & \multirow{2}{*}{30\%}& $\hat{N}_{RS}$ & 500.0 & 24.0 & 24.2 & 95.4 & 94.8 \\ 
  $N_{true}= 500 $ && $\hat{N}_{\hat{\psi}^{*}}$ & 499.6 & 19.6 & 23.9 & 98.1 (\textbf{94.8}) & 93.6 (\textbf{78.1}) \\ 
    & \multirow{2}{*}{50\%}& $\hat{N}_{RS}$ & 500.2 & 15.7 & 15.8 & 95.0 & 62.0 \\ 
  && $\hat{N}_{\hat{\psi}^{*}}$ & 500.1 & 12.7 & 15.8 & 98.4 (\textbf{95.1}) & 61.9 (\textbf{51.5}) \\ 
   \hline
\end{tabular}
\begin{tablenotes}
    \item[$a$] Wald-based CIs are evaluated for $\hat{N}_{RS}$ and $\hat{N}_{\hat{\psi}^*}$. The Bayesian credible intervals (from 10,000 Dirichlet posterior draws per simulation) for $\hat{N}_{\hat{\psi}^*}$ are reported in (\textbf{bold}).
    %\item[$b$] The standard error of $\hat{N}_{\hat{\hat{\psi}}^{*}}$ is calculated from a MI paradigm-based variance estimator (see Appendix A), instead of the delta method approach.
\end{tablenotes}
\end{threeparttable}
}
\end{table}

Table \ref{simu_results4} illustrates that estimation and inference based only on $\hat{N}_{RS}$ remain relatively stable after the finite population correction for all simulated scenarios, as expected. The ML estimator $\hat{N}_{\hat{\psi}^{*}}$ is uniformly more precise empirically, although its estimated variance becomes conservative as the prevalence and sampling rate into Stream 2 increase. Nevertheless, the proposed adjusted Bayesian credible interval (Appendix \ref{appendix_D}) performs well in terms of frequentist coverage. Importantly, across the full range of scenarios considered, this interval effectively translates the advantage of anchor stream-based CRC analysis in terms of estimation precision into inferential benefits as measured by interval width.

%%%%%%%%%%%%%%%%%%%%%%%%%%%%%%%%%%%%%%%%%%%%%%%%%%%%%%%%%%%%%%%%
%% section 4: real data example
\section{Data Analysis}

Cancer recurrence is a highly significant public health problem, and improvements in cancer survival rates can be made by screening and monitoring recurrence after the primary cancer diagnosis. CRISP (Cancer Recurrence Information and Surveillance Program) is developing a cancer recurrence surveillance database based on the Georgia Cancer Registry (GCR). The program aims to use up to six data streams to signal potential recurrences occurring among Georgia cancer survivors diagnosed with one of four cancer types (breast, prostate, colorectal, or lymphoma), and to generate data on the risks and rates of recurrence over ten years of follow-up. Through the longitudinal recurrence data, our plan is to leverage the proposed statistical methods to estimate the cancer recurrence case count using a CRC analysis framework.

Here, we apply our approach to CRISP cancer recurrence surveillance data to illustrate potential validity and precision improvements based on the proposed estimators. Our goal is to estimate the total number of breast cancer recurrent patients from the target population among registry patients who were diagnosed with breast cancer during 2013-2015 while living in one of the 5 metro Atlanta counties, with their initial cancer directed surgery completed at one of five facilities with Commission on Cancer (CoC) accreditation in this same region. Hospitals participating in the Commission on Cancer Program in Georgia report data to the GCR on cancer recurrences among patients diagnosed and/or treated at their facility. The CoC data stream is potentially non-representative of the target population, while also yielding some false positive recurrence status reports from the hospitals relative to the CRISP protocol-based recurrence definition. In this motivating example, we thus use CoC recurrence data as the existing error-prone surveillance data stream, or Stream 1. Subsequently, on May 1st 2022, an actual anchor stream sample was drawn from the target breast cancer patient population to serve as Stream 2 for the two-stream CRC analysis. Our goal is to estimate the number of recurrences that have occurred among the target population as of that date.

In this example, the size of the target population is 1,029. Among these breast cancer patients, there were 83 recurrent records documented in Stream 1. For the anchor stream, a random sample of 200 patients (19.4\% of the population) was drawn from the target population, with 31 patients captured as true recurrent signals. All randomly sampled patients in the anchor stream were reviewed based on CRISP protocol-based medical report abstraction, which we treat as a gold-standard diagnostic test for recurrence. Table \ref{Obs_type_realdata} shows the summary of capture profiles from these two data streams. Table \ref{Obs_type_realdata2} (a realization of Table \ref{collapse_table_with_prob}) shows the detailed cell counts of each observation type from these two data streams.

% \begin{table}[ht]
%     \centering
%     \caption{Capture-Recapture Table for Real Data Example}
%     \label{Obs_type_realdata}
%     \begin{tabular}{|c|ccc|c||c|ccc|c|}
%     \hline
%     \multicolumn{5}{|c||}{$A$. Validated Records} & \multicolumn{5}{c|}{$B$. Original Records }\\
%     \multicolumn{5}{|c||}{(Accurate Disease Signals)} & \multicolumn{5}{c|}{(w/ misclassification)}\\
%     \hline
%     \# of Cases &\multicolumn{4}{c||}{ Found in Stream 2}  & \# of Cases &\multicolumn{4}{c|}{Found in Stream 2}\\
%     \hline
%      Found in Stream 1 && Yes && No & Found in Stream 1 && Yes && No\\
%     \hline
%      Yes && 12  &&  50 &Yes &&  12 &&  62\\
%      \hline
%      No\tnote{1}  &&  6 &&  ? &No  &&  6 &&  ? \\
%      \hline
%     \end{tabular}
% \end{table}

\begin{table}[ht]
    \centering
    \caption{Capture-Recapture Table for Real Data Example}    \label{Obs_type_realdata}
    \begin{threeparttable}[b]
    \begin{tabular}{|c|c|c|c|}
    \hline
     &\multicolumn{2}{c|}{\# of Cases Found in Stream 2}  &  \\
    \hline
     \# of Cases Found in Stream 1 & \hspace*{10mm} Yes \hspace*{10mm} &\hspace*{10mm}  No  \hspace*{10mm} & \hspace*{2mm} Total \hspace*{2mm}\\
    \hline
     Yes & 14 & 69 & 83\\
     \hline
     No   &   17 & ? &  \\
     \hline
     Total & 31 &   & $N = ?$\\
     \hline
    \end{tabular}
    \end{threeparttable}
\end{table}

\begin{table}[ht]
\centering
\caption{Cell Counts of Observation Table in Real Data Example}
\label{Obs_type_realdata2}
  \begin{tabular}{|c|l|c|c|} 
 \hline
  Cell & \multicolumn{1}{c|}{Observation Type} & Original Counts  \\ 
 \hline
 \multirow{2}{*}{$n_1$} & Test + in Stream 2, &\multirow{2}{*}{14}  \\
  & Test + in Stream 1  &  \\ \hline
  \multirow{2}{*}{$n_2$} & Test + in Stream 2, & \multirow{2}{*}{17} \\\
  & Not test + in Stream 1 &  \\\hline
  \multirow{2}{*}{$n_3$} & Test \textminus ~ in Stream 2, &\multirow{2}{*}{3}  \\
  & Test + in Stream 1  &  \\ \hline
  \multirow{2}{*}{$n_4$} & Test \textminus~in Stream 2, & \multirow{2}{*}{166} \\
  & Not test + in Stream 1  & \\ \hline
  \multirow{2}{*}{$n_5$} & Not sampled in Stream 2, &\multirow{2}{*}{66}  \\
  & Test + in Stream 1  & \\ \hline
   \multirow{2}{*}{$n_6$} &  Not sampled in Stream 2, & \multirow{2}{*}{763} \\
  & Not test + in Stream 1  & \\
  \hline
 \end{tabular}
\end{table}

Results utilizing the estimators discussed in the Methods section are shown in Table \ref{realdata_estimation}. For the anchor stream data, we apply the existing sampling-based method to estimate the recurrent case count as an unbiased estimation benchmark. For the CRC analysis, we incorporate the classic Chapman estimator to illustrate the naive estimation without justifying any misclassification. Then to account for error-prone diagnostic status in Stream 1, we use our proposed ML estimator ($\hat{N}_{\hat{\psi}^{*}}$) in equation (\ref{MLE_collapse}) to estimate the total case count. In this real data example, our only source of information about the $PPV_1$ parameter derives from the anchor stream design; hence, we do not report the proposed plug-in estimate from equation (\ref{MLE_psi}).

\begin{table}[ht]
\centering
\caption{Estimation Results in Real Data Example}
\label{realdata_estimation}
\begin{threeparttable}[b]
 \begin{tabular}{lcccl} 
 \hline
  Estimator & $\hat{N}$ & SE & 95\% CI & Note\\ 
  \hline
  $\hat{N}_{RS}$ & 159.5 & 23.7 & [113.1, 205.9] \tnote{$a$} & Uses Stream 2 Only\\
  $\hat{N}_{Chap}$ & 178.2 & 29.6 & [138.5, 279.8]\tnote{$b$} & Naive CRC Estimation\\
  %$\hat{N}_{{\psi}}$ & 164.1 & 19.3 & [126.2, 201.9]\tnote{$a$}  & Use $PPV_1 = 91.78$\%\\
  $\hat{N}_{\hat{\psi}^{*}}$ & 156.2 & 20.7 & [115.7, 201.9]\tnote{$a$},~ \textbf{[118.5, 198.8]}\tnote{$c$} & Uses $\hat{PPV}_1 = 82.35$\%\\
  \hline
 \end{tabular}
    \begin{tablenotes}
        \item[$a$] Wald CI
        \item[$b$] Transformed logit CI \cite{Sadinle2009}
        \item[$c$] Unadjusted Bayesian Credible Interval (see Appendix \ref{appendix_D})
  \end{tablenotes}
\end{threeparttable}
\end{table}

Table \ref{realdata_estimation} shows the estimation results for the breast cancer recurrence case count from the observed surveillance data streams. First, we consider the RS-based estimator ($\hat{N}_{RS}$) as an unbiased standard estimate for the total count. Based on Stream 2 only, the estimated number of breast cancer recurrences among the registry-based target population is 159.5 with 95\% CI [113.1, 205.9]. This estimate lacks precision due to small sample size. Therefore, we seek to incorporate additional information in Stream 1 to improve the estimation efficiency via CRC analysis. 

Evidently, the naive classical Chapman estimate ($\hat{N}_{Chap}$) reflects potential upward bias caused by the misclassification effect (i.e., false positives in Stream 1), and the accompanying transformed logit-based confidence interval is extremely wide. In contrast, the proposed anchor stream-based estimate ($\hat{N}_{\hat{\psi}^{*}}$) agrees well with $\hat{N}_{RS}$, while reflecting improved precision through its use of the Stream 1 data. This estimator incorporates bias adjustment through the estimable positive predictive value parameter ($\hat{PPV}_1 = n_1/(n_1+n_3) = 82.35\%$) characterizing the CoC recurrence designations in Stream 1. The preferred estimated total case count is thus 156.2 with 95\% CI [118.5, 198.8] based on the proposed adjusted Bayesian credible interval. The cumulative proportion of breast cancer recurrences is estimated as $\hat{p} = \hat{N}_{\hat{\psi}^{*}}/N_{tot} = 15.2\%$, with 95\% CI [11.5\%, 19.3\%]. To our knowledge, these are the first principled empirical estimates of the breast cancer recurrence case count and cumulative incidence rate among metro Atlanta-based Georgia Cancer Registry patients.

%While the validation data might not always be available for analysis in practice, then the proposed ML estimator $\hat{N}_{\hat{\psi}^*}$ is the most reliable estimator with justifying the uncertainty from the positive predictive value in the existing surveillance data stream. In this example, the unbiased estimate from the random sample estimator also falls into this 95\% credible interval. Therefore, we report the breast cancer recurrence case count from our target population (breast cancer patients from the 5 metro Atlanta counties) as 137.8 [97.4, 197.9] and the prevalence as 12.9\% [9.1\%, 18.5\%]. To the best of our knowledge, these are the first principled, empirical estimates of the prevalence of breast cancer recurrence rate in the Metro Atlanta area.

%%%%%%%%%%%%%%%%%%%%%%%%%%%%%%%%%%%%%%%%%%%%%%%%%%%%%%%%%%%%%%%%
%% section 5: Discussion and Conclusion
\section{Discussion}

%. conclusion
In this article, anchor stream-based CRC estimators are proposed to estimate epidemiological disease case counts while accounting for misclassified diagnostic disease status. This design-based approach assures the independence assumption utilized in the classical two-stream CRC approach, and applies “gold-standard” or perfect diagnostic testing for individuals selected into what can be a relatively small representative sample. Our empirical studies suggest that the proposed estimators provide unbiased and efficient estimation for total case counts, with enhanced precision relative to the sampling-based estimator alone. A real data example focused on estimation of the GCR-based breast cancer recurrent case counts in the metro Atlanta area illustrates the direct application of the proposed estimators.

When dealing with a closed registry population from which a representative sample can be drawn and true disease status ascertained by a method such as protocol-based records abstraction (as in the CRISP motivating example), the proposed method for anchor stream-based CRC analysis becomes highly appealing in practice. It only requires the documentation of positive signals in the non-anchor surveillance stream, and permits valid estimation of the true case count based on an estimable positive predictive value (PPV) parameter. Our simulation studies demonstrate that collecting even a small anchor stream sample, to augment what might be a much larger but highly non-representative and error-prone existing surveillance data stream, can unlock a far superior estimator than can be derived based on classical CRC analysis or the anchor stream sample alone. Moreover, extending to a multi-stream surveillance setting, we noted that a combination strategy is immediately available by pooling all non-representative data streams together into one “composite” stream. The proposed ML estimator of the case count adjusting for the non-representativeness and misclassification in the non-anchor streams can then be directly applied, treating the composite stream as Stream 1.

We believe the proposed estimators could provide new perspectives for epidemiological surveillance studies in terms of both study design and analysis, especially when faced with misclassified disease signals based on existing surveillance systems. The key is an enumerable registry of members of the target population, and the ability to randomly sample from that registry and determine true case status for those selected. In addition to monitoring cumulatively incident cases as in CRISP, other potential applications of the design and proposed statistical approach include the periodic monitoring of infectious diseases (e.g., COVID-19) in closed community settings. In this direction, we aim in future work to adapt the approach toward adjusting the error-prone counts through the utilization of manufacturer-specified diagnostic test sensitivity and specificity parameters. This extension would also allow for the use of an imperfect diagnostic test in the anchor stream sample, thereby further expanding the practical uses of the design and analytic methods introduced here.

\vspace{2cm}
% BIBLIOGRAPHY
\addcontentsline{toc}{chapter}{Bibliography}
\bibliographystyle{apalike}
%\bibliography{references}
%\bibliographystyle{ownbib}

\newpage
\appendix
\numberwithin{equation}{section}

\section{Variance Calculation from Multiple Imputation Paradigm} \label{appendix_A}

We propose an approach to estimating the variances of the proposed anchor stream-based estimators proposed in this article based on a multiple imputation paradigm \cite{Rubin1987} considering $\hat{PPV_1}(m_{11}+m_{10})$ as a missing value to propagate the uncertainty caused by misclassification in Stream 1. 

Suppose we have $m>1$ independent imputations of $\hat{PPV_1}(m_{11}+m_{10})$, thereby obtaining $m$ estimates $\hat{N}_{{\psi}}^{(1)}, \cdots, \hat{N}_{{\psi}}^{(m)}$. The corresponding within-imputation variance estimates $\hat{U}^{(1)}, \cdots, \hat{U}^{(m)}$ are evaluated via the original anchor stream variance estimator $\hat{U} = m_{01}(1-{\psi})/{\psi}^2$ in Lyles et al. (2022) \cite{Lyles2022}. Then the multiple imputation (MI) estimator for $\hat{N}_{{\psi}}$ is calculated by taking the average of all imputations as $\hat{N}_{{\psi}, MI} = m^{-1} \sum \hat{N}_{{\psi}}^{(j)}$ and the total variance of $\hat{N}_{{\psi}, MI}$ includes two components, the within-imputation variance $\bar{U} = m^{-1}\sum \hat{U}^{(j)}$ and between-imputation variance $B = (m-1)^{-1}\sum (\hat{N}_{{\psi}}^{(j)} - \hat{N}_{{\psi}, MI})^2$. Hence the estimated total variance from the multiple imputation paradigm is as follows. 
\begin{equation}\label{MIvar}
    \hat{Var}(\hat{N}_{{\psi}, MI})= (1+m^{-1})B+\bar{U}.
\end{equation}

Intuitively, the within-imputation component $\hat{U}^{(m)}$ in (\ref{MIvar}) is invariant across imputations. As for the between-imputation component, we have two versions of the variance calculations, as follows: 

\begin{itemize}
    \item When the parameter for $PPV_1$ is assumed known or is very precisely estimated by means of extensive validation data, the missing value can be imputed $M$ times via independent binomial draws to obtain the number of successes among  $(m_{11} + m_{10})$ subjects based on the fixed probability $PPV_1$. 
    \item In the typical case of an unknown positive predictive value $PPV_1$, each imputation is conducted using a two-stage procedure. For the $m$th imputation, we start with a weakly-informative Jeffreys’ $Dirichlet$(0.5, 0.5, 0.5) prior for the probabilities $p_1$, $p_3$ and $p_5$ corresponding to the cell counts $n_1$, $n_3$ and $n_5$ in Table \ref{collapse_table_with_prob}. We obtain the estimate $\Tilde{PPV}_1^{(m)}$ by inserting the resulting posterior draws (see \ref{dirichlet_AppendixD}) into the expression for the MLE for $PPV_1$. Then the binomial imputation is implemented as in the first case above, based on the probability  $\Tilde{PPV}_1^{(m)}$.
    %\item As for the situation $PPV_1$ is unknown, another two-stage imputation idea is implemented to show better performance. Instead of imputing from $\Tilde{PPV}_1^{(m)}(m_{11} + m_{10})$, the empirical study shows that the imputation $\Tilde{PPV}_{10}^{(m)}m_{10}$ is better in matching the empirical standard derivation of the MLE estimator.
\end{itemize}

\section{Derivation for Closed-form MLE for \texorpdfstring{$N$}{Lg}} \label{appendix_B}

The prevalence estimator is derived using the following probability chain:
\begin{align*}
    Pr(true~+)  =& Pr(true~+,~sampled~in~1)+Pr(true+,~not~sampled~in~1)\\
     =& Pr(true~ +|~sampled~in~1,~test+)Pr(test~+|~sampled~in~1)Pr(sampled~in~1)+\\
    & Pr(true~ +|~sampled~in~1,~test-)Pr(test~-|~sampled~in~1)Pr(sampled~in~1)+\\
    & Pr(true~+|~not~sampled~in~1)Pr(not~sampled~in~1)\\
    =& PPV_1 \pi^*_{s1}\phi + (1-NPV_1)(1-\pi^*_{s1})\phi + \pi_{\bar{s}1}(1-\phi) \\%~~~~~~~~~~~ \textit{(full table)} \\
    =& PPV_1\theta + \pi -\theta % ~~~~~~~~~~~~~~~~~~~~~~~~~~~~~~~~~~~~~~~~~~~~~~~~~~~~~~~~~~~~~~~ \textit{(collapse table)}
\end{align*}

Hence, the estimator $\hat{N}_{\hat{\psi}^{*}}$ of the total case number $N$ is calculated via the following formula, and this is algebraically equivalent to ML estimator in (\ref{MLE_collapse}):
\begin{align}
    \hat{N}_{\hat{\psi}^{*}} = N_{tot}\hat{Pr}(true+) 
    =  N_{tot}(\hat{PPV}_1\hat{\theta} + \hat{\pi} -\hat{\theta}).
\end{align}

The variance estimator for $\hat{N}_{\hat{\psi}^{*}}$ is calculated using the version of the MI paradigm that treats $PPV_1$ as unknown and replaces $\psi$ by $\hat{\psi}^*$ (see Appendix \ref{appendix_A}).

%The variance estimator for $\hat{N}_{\hat{\psi}^{*}}$ is derived based on the multivariate delta method. 

%Let $g(\Theta) = N_{tot}({PPV}_1{\theta} + {\pi} -{\theta})$, where $\Theta = [\psi, \theta, \pi, PPV_1]^T$. Then $g'(\Theta) = N_{tot}[0, PPV_1-1, 1, \theta]^T$. From the likelihood function, we have
%\begin{align*}
%   \Sigma_{\Theta} = \begin{bmatrix}
%\frac{n_1+n_2+n_3+n_4}{\psi^2}+\frac{n_5+n_6}{(1-\psi)^2} & 0 & 0 & 0 \\
%0 & \frac{n_1+n_3+n_5}{\theta^2}+\frac{n_4}{(\pi-\theta)^2} + \frac{n_6}{(1-\theta)^2} & -\frac{n_4}{(\pi-\theta)^2} & 0 \\
%0 & -\frac{n_4}{(\pi-\theta)^2} & \frac{n_2}{(1-\pi)^2}+\frac{n_4}{(\pi-\theta)^2} & 0 \\
%0 & 0 & 0 & \frac{n_1}{PPV_1^2} + \frac{n_3}{(1-PPV_1)^2}
%\end{bmatrix} ^{-1}
%\end{align*}
%and then the variance from the multivariate delta method is calculated as
%\begin{align}
%    \hat{Var}(\hat{N}_{\hat{\psi}^{*}}) = g'(\Theta)^T\Sigma_{\Theta} g'(\Theta)|_{\Theta = \hat{\Theta}}
%\end{align}

\section{Proof of the Relationship between \texorpdfstring{$PPV_1$}{Lg} and \texorpdfstring{$PPV_{10}$}{Lg}} \label{appendix_C}

Assuming we conduct a two-stream CRC analysis with Stream 2 as anchor stream.\par

Firstly, we define the following probability event sets:

\begin{itemize}
    \item $A = \{true ~+\}$
    \item $B = \{sampled~in~1 ~\cap ~ test~+ ~\cap ~sampled~in~2 ~\cap~ test~-\}$
    \item $C = \{sampled~in~1 ~\cap ~ test~+ ~\cap ~not~sampled~in~2\}$
    \item $D = \{sampled~in~1 ~\cap ~ test~+ ~\cap ~sampled~in~2 ~\cap~ test~+\}$
\end{itemize}

Intuitively, we have the union of sets among $B$, $C$ and $D$ as $(B\cup C\cup D) = \{sampled~in~1~\cap~test~+\}$. Next, for the definitions of $PPV_1$ and $PPV_{10}$, we have the following derivations:
\begin{align*}
    PPV_1 &= Pr(true~+ |~sampled~in~1~\cap~test~+)\\
    &= \frac{Pr(true~ + ~\cap~ sampled~in~1 ~\cap~ test ~+)}{Pr(sampled ~in~1 ~\cap~ test~ +)} \\
    &= \frac{Pr(A\cap (B\cup C\cup D))}{Pr(B\cup C\cup D)} \\
    &= \frac{Pr[(A\cap B) \cup (A\cap C) \cup (A\cap D)]}{Pr(B) + Pr(C) + Pr(D)} \\
    & = \frac{Pr(A\cap B) + Pr(A\cap C) + Pr(D)}{Pr(B) + Pr(C) + Pr(D)} \\
    PPV_{10} & = Pr(true~+ |~(sampled~in~1~\cap~test~+) \cap~[(sampled~in~2~\cap~test~-)\cup(not~sampled~in~2)]) \\
    & = \frac{Pr[(A\cap B) \cup (A\cap C)]}{Pr(B\cup C)} \\
    & = \frac{Pr(A\cap B) + Pr(A\cap C) }{Pr(B) + Pr(C)} 
\end{align*}

Therefore, the relationship between $PPV_1$ and $PPV_{10}$ is as follows.
\begin{align} \label{PPV_10}
    PPV_{10} &= [1+\frac{Pr(D)}{Pr(B) + Pr(C)}]PPV_1 - \frac{Pr(D)}{Pr(B) + Pr(C)} \nonumber\\
    &= [1+ \frac{p_1}{p_3+p_5}] PPV_1 - \frac{p_1}{p_3+p_5}
\end{align}

The MLE for $PPV_{10}$ is obtained by replacing $p_1$, $p_3$ and $p_5$ by $n_1$, $n_3$ and $n_5$ in Table \ref{collapse_table_with_prob} and inserting the MLE for $PPV_1$.

%where $n_1$, $n_3$ and $n_5$ are defined in table \ref{collapse_table_with_prob}. When the MLE $\hat{PPV}_1$ is included into \ref{PPV_10}, we could get the MLE for $PPV_{10}$ as well.
%\begin{align} \label{PPV_10_hat}
%    \hat{PPV}_{10} &= [1+ \frac{n_1}{n_3+n_5}] \hat{PPV}_1 - \frac{n_1}{n_3+n_5}  \nonumber\\
%    &= [1+ \frac{n_1}{n_3+n_5}] \frac{n_1}{n_1+n_3} - \frac{n_1}{n_3+n_5} \nonumber\\
%    & = \frac{n_1n_5}{(n_1+n_3)(n_3+n_5)}.
%\end{align}

\section{Details for Calculating Proposed Bayesian Credible Intervals}\label{appendix_D}

In the first step we start with quantifying the uncertainty for $\hat{PPV}_{10}$ based on a weakly-informative Jeffreys’ $Dirichlet$(0.5,0.5,0.5) prior for the probabilities $p_1,~p_3,~p_5$ corresponding to the cell counts $n_1,~n_3,~n_5$ in Table \ref{collapse_table_with_prob}.  This prior yields the following posterior distribution for $\bm p^{*} = (p_1^*,~p_3^*,~p_5^*)$ given the observed cell counts $\bm n = (n_1,n_3,n_5)$.
\begin{align}\label{dirichlet_AppendixD}
  \bm{p}^{*}|\bm{n} = (n_1,n_3,n_5) \sim Dirichlet (n_1+0.5, n_3+0.5, n_5+0.5).
\end{align}
For each posterior draw from this distribution, the posterior positive predictive value $\Tilde{PPV}_{10,s}^*, (s = 1, \cdots, S)$ is calculated via (\ref{PPV_10}) in Appendix \ref{appendix_C} and then the false positive case counts in $m_{10}$ are adjusted by means of this positive predictive value, i.e., $\hat{m}^*_{10,s} = m_{10}\Tilde{PPV}_{10,s}^*$.

In the next step, we continue to explain the natural uncertainty in the estimated case count $N$. In conjunction with each posterior draw $\hat{m}^*_{10,s}$ ($s = 1,2,\cdots S$), we subsequently draw posterior samples for the three multinomial proportions $\bm p^{*}_{s} = (p_{11,s}^{*},p_{10,s}^{*},p_{01,s}^{*})$ based on a second independent Jeffreys’ $Dirichlet$(0.5,0.5,0.5) prior, where $p_{ij,s}^{*}$ represents the posterior probability of the accurate cell count $m_{ij}$ in Table \ref{Obs_type0}, conditional on being captured by at lease one of the two data sources and assuming no misclassification. The conjugate prior yields the following posterior distribution for $\bm{p}^{*}_{s}$ given $\bm {m}_{s}^* = (m_{11},m_{10,s}^*,m_{01})$, where $m_{11} = n_1$ and $m_{01} = n_2$ from Table \ref{collapse_table_with_prob}.
%Meanwhile, we have $\bm {n}_{m}^* = (\hat{n}_{11},\hat{n}_{10,m}^*,\hat{n}_{01}) = (n_1, \hat{n}_{10,m}^*, [n_4+n_7])$ and hence $\hat{n}_{c,m} =  n_1 + \hat{n}^*_{10,m} + (n_4+n_7)$. 
\begin{align}\label{BC_distr}
    \bm{p}^{*}_{s}|\bm{m}_{s}^* \sim Dirichlet (m_{11}+0.5, m_{10,m}^*+0.5, m_{01}+0.5)
\end{align}

For each posterior sample $\bm p_{s,t}^* = (p_{11,(s,t)}^{*},p_{10,(s,t)}^{*},p_{01,(s,t)}^{*})$ $(t = 1,\cdots, T)$ from (\ref{BC_distr}), we calculate the unconditional posterior probability of capture in stream 1 as $\Tilde{p}_{1,(s,t)}^{*} = \hat{\psi}^{*}(p^{*}_{11,(s,t)}+p^{*}_{10,(s,t)})[\hat{\psi}^{*}(p^{*}_{11,(s,t)}+p^{*}_{10,(s,t)}) + p^{*}_{01,(s,t)}]^{-1}$, from which we further calculate the unconditional posterior probabilities $\Tilde{\bm p}^{*}_{s,t} = (\Tilde{p}^{*}_{11,(s,t)},\Tilde{p}^{*}_{10,(s,t)},\Tilde{p}^{*}_{01,(s,t)})$ as follows \cite{Lyles2022}.
\begin{align}
    \Tilde{p}_{11,(s,t)}^{*} &= \Tilde{p}_{1,(s,t)}^{*}p^{*}_{11,(s,t)}(p^{*}_{11,(s,t)}+p^{*}_{10,(s,t)})^{-1} \nonumber\\
    \Tilde{p}_{10,(s,t)}^{*} &= \Tilde{p}_{1,(s,t)}^{*}p^{*}_{10,(s,t)}(p^{*}_{11,(s,t)}+p^{*}_{10,(s,t)})^{-1} \nonumber\\
    \Tilde{p}_{01,(s,t)}^{*} &= \hat{\psi}^{*}(1-\Tilde{p}_{1,(s,t)}^{*})  \nonumber
\end{align}
where $\hat{\psi}^{*}$ is defined in the expression for the ML estimator in  (\ref{MLE_collapse}). 

Conditional on $m_{c,s}^* = m_{11} + m^*_{10,s} + m_{01}$, the posterior draw of $N$ is calculated as $\hat{N}_{\hat{\psi}^{*},(s,t)|m_{c,s}^*}^* = m_{c,s}^* / \Tilde{p}_{c,(s,t)}^*$, where $\Tilde{p}_{c,(s,t)}^* = \Tilde{p}_{11,(s,t)}^{*} + \Tilde{p}_{10,(s,t)}^{*} +\Tilde{p}_{01,(s,t)}^{*}$. However, the variance of $\hat{N}_{\hat{\psi}^{*},(s,t)|m_{c,s}^*}^*$ is overly optimistic for configuring a reliable Bayesian credible interval \cite{Lyles2022}, due to conditioning on $m_{c,m}^*$. To account for this we generate a new total number captured, denoted $\Tilde{m}_{c,(s,t)}^*$ from a binomial distribution with size $\hat{N}_{\hat{\psi},(s,t)|m_{c,s}^*}^*$ and sampling probability $\Tilde{p}_{c,(s,t)}^*$. The posterior draw for the vector of the cell counts $\Tilde{\bm m}_{s,t}^* = (\Tilde{m}_{11,(s,t)}^*, \Tilde{m}_{10,(s,t)}^*, \Tilde{m}_{01,(s,t)}^*)$ is calculated as $\Tilde{m}_{c,(s,t)}^*\Tilde{\bm p}^{*}_{s,t}$. Then the posterior estimate for total case number $\hat{N}_{\hat{\psi}^*,(s,t)}^*$ is updated by 
\begin{equation}\label{BC_unadj}
    \hat{N}_{\hat{\psi}^*,(s,t)}^* = \Tilde{m}_{11,(s,t)} + \Tilde{m}_{10,(s,t)} + \Tilde{m}_{01,(s,t)}/\hat{\psi}^{*}.
\end{equation} 

We consider two adapted Bayesian credible intervals, with the choice between them dictated by the estimated disease prevalence. First, we define the unadjusted credible interval, $(LL_{unadj}, UL_{unadj})$, as the (2.5\%, 97.5\%) percentile interval from all $S\times T$ posterior estimates in (\ref{BC_unadj}). This is preferable in the case of a small estimated prevalence; we suggest its use based on a 20\% threshold for the estimated prevalence (i.e., when $\hat{p} = \hat{N}_{\hat{\psi}^*}/N_{tot} < 20\%$). 

When the prevalence is estimated over 20\%, an adjusted credible interval, $(LL_{adj}, UL_{adj})$, is designed to incorporate a scale and shift adjustment. These adjustments are similar to those proposed in conjunction with the original anchor stream estimator \cite{Lyles2022}, with some updates to account for misclassification. Continuing from (\ref{BC_unadj}), we propose to scale and shift each draw $\hat{N}_{\hat{\psi}^*,(s,t)}^*$ by the following calculation.
\begin{equation}
   \mathbb{\hat{N}}_{s,t} = a\hat{N}_{\hat{\psi}^*,(s,t)}^* + b,
\end{equation}
where $a = \sqrt{\hat{Var}(\hat{N}_{comp, B})/\hat{Var}(\hat{N}_{\hat{\psi},MI})}$, $b = \hat{N}_{\hat{\psi}^*}(1-a)$ and $\hat{Var}(\hat{N}_{comp, B})$ is an altered version of a closed-form variance estimator proposed in Lyles et al. (2022) \cite{Lyles2022}. 
\begin{equation}\label{BC_adj1}
    \hat{Var}(\hat{N}_{comp, B}) = [\hat{Var}(\hat{N}_{RS})^{-1} + \hat{Var}(\hat{N}_{LP,MI})^{-1}]^{-1}
\end{equation}

The alteration in (\ref{BC_adj1}) is reflected in the second term within the brackets, where we use a multiple imputation-based version of the variance of the classic Lincoln-Petersen estimator \cite{Seber1982}, i.e.,  $\hat{Var}(\hat{N}_{LP}) = \frac{(n_{11} + n_{10})(n_{11}+n_{01})n_{10}n_{01}}{n_{11}^3}$. We calculate $\hat{Var}(\hat{N}_{LP})$ by inserting PPV adjusted cell counts for $(n_{11}, n_{10}, n_{01})$ and accounting for uncertainty using the the same imputation paradigm \cite{Rubin1987} described in Appendix \ref{appendix_A}. 

We denote the credible interval based on the (2.5\%, 97.5\%) percentile of the random posterior draws $\mathbb{\hat{N}}_{s,t}$, as $(LL_{ab}, UL_{ab})$. Meanwhile, we define the lower and upper limits of the FPC-adjusted Ward-type CI accompanying $\hat{N}_{RS}$ as $(LL_{RS}, UL_{RS})$. We utilize these to temper potential improvement in CI width based on $(LL_{ab}, UL_{ab})$ against the risk of overcoverage. The adjusted credible interval, denoted $(LL_{adj}, UL_{adj})$, is defined by the following limits:
\begin{equation}
    LL_{adj} = \max\{LL_{ab}, \frac{LL_{ab} + LL_{RS}}{2}\}, ~~UL_{adj} = \min\{UL_{ab}, \frac{UL_{ab} + UL_{RS}}{2}\}
\end{equation}

\end{document}